\documentclass{article}
\usepackage{amsmath}
\usepackage{amssymb}
\usepackage[T1]{fontenc}
\usepackage[utf8]{inputenc}
\usepackage{lmodern}
\input{tcilatex}
\begin{document}

\noindent {\large Light-Cones, Almost Light-Cones and Almost-Complex
Light-Cones }

Ezra T. Newman,

Dept. of Physics and Astronomy, Univ. of Pittsburgh

(4/6/17)

\textbf{Abstract}

We point out (and then apply to a general situation) an unusual relationship
among a variety of null geodesic congruences; (a) the generators of \textit{%
ordinary light-cones and (b) certain (related) shear-free but twisting
congruences in Minkowski Space-time as well as (c) asymptotically shear-free}
null geodesic congruences that exist in the neighborhood of Penrose's $%
\mathcal{I}^{+}$\ in Einstein or Einstein-Maxwell asymptotically
flat-space-times. We refer to these geodesic congruences respectively as: 
\textit{Lignt-Cones (LCs),\ as \textquotedblleft Almost-Complex"-
Light-Cones,} (ACLCs), [though they are real they resemble complex
light-cones in complex Minkowski space]\ and finally to a family of
congruences in asymptotically flat-spaces as `\textit{Almost Light-Cones}',
(ALC). \ The two essential points of resemblance among the three families
are: (1) they are all either shear-free or asymptotically shear-free and (2)
in each family the individual members of the family can be labeled by the
points in a real or complex four-dimensional manifold. \ As an example, the
Minkowski space LCs are labeled by the (real) coordinate value of their
apex. In the case of (ACLCs) (complex coordinate values), the congruences
will have non-vanishing twist whose magnitude is determined by the imaginary
part of the complex coordinate values.

In studies of gravitational radiation, Bondi-type of null surfaces and their
associated Bondi coordinates have been almost exclusively used for
calculations. It turns out that some surprising relations arise if, instead
of the Bondi coordinates, one uses ALCs and their associated coordinate
systems in the analysis of the Einstein-Maxwell equations in the
neighborhood of $\mathcal{I}^{+}$. More explicitly and surprisingly, the
asymptotic Bianchi Identities (arising directly from the Einstein
equations), expressed in the coordinates of the ALCs, turn directly into
many of the standard definitions and equations and relations of classical
mechanics coupled with Maxwell's equations. \ These results extend\ and
generalize the beautiful results of Bondi and Sachs with their expressions
for, and loss of, mass and linear momentum.

\section{Introduction}

It has been known for many years that \textit{shear-free null geodesic
congruences}, in flat, Einstein and Einstein-Maxwell spaces have fascinating
and useful properties. \ The Robinson -Trautman metrics and the
Goldberg-Sachs theorem are among the most prominent examples.

It is one of our purposes to explore some new aspects of these congruences.
\ We first note that the generators (the null geodesics) of ordinary
light-cones, LCs, in Minkowski space-time are among the most familiar
examples of shear-free congruences. \ Each of these can be identified by the
four Minkowski coordinate values at their apex. \ We will show, in Sec.II,
that another set of SFGCs, can be described and identified by four complex
valued coordinates that can be thought of as coordinates in \textit{complex}
Minkowski space-time. Although these congruences are \textit{real
congruences in real Minkowski space}, we will refer to them as
Almost-Complex Light-Cones, (ACLCs), because of their identification labels
- and their close relation to complex LCs cones. \ The real LCs are special
cases of the ACLCs. The two sets, LCs and ACLCs, constitute the entire class
of flat-space shear-free congruences that are diverging with isolated
caustics.

A second type of related null geodesic congruence, described in Sec.III,
occur in asymptotically flat-space-times - where, though they are \textit{not%
} shear-free, are instead \textit{asymptotically shear-free}. The individual
members, as in the flat-space-time, are also labeled by four complex numbers
and are referred to as Almost-Light-Cones, ALCs. \ The ALCs contain, as
special cases, all the other cases. The four-complex numbers (in each of the
cases) define a complex manifold, referred to as $\mathcal{H}$-space,
containing an interesting variety of properties, e.g., a complex metric that
satisfies the complex vacuum Einstein equations with a self-dual Weyl tensor.

In the cases when the labels are real the congruences are surface forming, -
when they are complex, the imaginary parts are a measure of the twist of the
congruence.

\textbf{Note\ }We have taken the liberty to slightly generalize the meaning
of "congruence". Usually it refers to a three parameter family of curves
filling a space-time region. We will use it to mean a two parameter family
of curves, as, for example, the geodesics on \textit{one single light cone}.
\ A one-parameter family of such light-cones (based on a time-like world
line) would yield the standard example of a congruence.

Since the 1950s most studies of the far-field gravitational properties,
including gravitational radiation, have used, as technical tools in the
analysis of the Einstein or Einstein-Maxwell equations, certain null
surfaces (Bondi surfaces) and the associated coordinate systems referred to
as Bondi coordinates. \ This led to Bondi's and Sach's, beautiful theorems
on mass and linear\ momentum loss\cite{Bondi}\cite{Sachs} and eventually to
the development of LIGO with its technology. \ Finally, after over 50 years
of developments involving theory, numerical analysis and observational work,
this led to the observation and analytic understanding of the collision and
merger of the pair of black holes that produced the gravitational wave
signal, GW105,\cite{GW}, that was seen by LIGO early in 2016.

It has seemed for many years as if the Bondi system was almost sacrosanct -
the best and virtually only way to study, in general,\ the asymptotic
behavior of the Einstein-Maxwell equations. Our contention is that this is
not necessarily so - there appear to be very good reasons to consider the
use of families of ALCs as our choice of asymptotic coordinate systems. \
First of all they \textit{very closely resemble} the standard Minkowski
space null coordinate systems in the neighborhood of null infinity; they are
labeled by four coordinates and are asymptotically-shear-free - the Bondi
surfaces are not. Second, when the asymptotic Bianchi Identities are studied
in families of ALCs they turn out to explicitly be - in the low order, ($%
l=0,1$) spherical harmonic decomposition - many of the standard definitions
and dynamic relations and equations of classical mechanics coupled with the
Maxwell field. They are a large extension of the Bondi-Sachs results. Though
this itself is surprising, the astonishing and so-far inexplicable fact is
that these equations of classical mechanics take place in the $\mathcal{H}$%
-space rather than in physical space-time. Sec. IV will contain a
description of these results. \ 

\subsection{Flat-Space Shear-Free Congruences}

Using $x^{a}$ as standard Minkowski-space coordinates, an arbitrary null
geodesic congruence can be described by

\begin{eqnarray}
x^{a} &=&\sqrt{2}ut^{a}-L\overline{m}^{a}-\overline{L}m^{a}+(r-r_{0})l^{a},
\label{ngc} \\
\sqrt{2}t^{a} &=&l^{a}+n^{a} \\
u_{r} &=&\sqrt{2}u,  \notag
\end{eqnarray}%
where the `parameters' ($u,\zeta ,\overline{\zeta }$) label the individual
members of the geodesic congruence and $r$ is the affine parameter along
each null geodesic, $u_{r}=t-r,\ $is the retarded time. The$\ L(u,\zeta ,%
\overline{\zeta })\ $(which is a null angle field\cite{PandR} and the
primary source of information about the congruence) is an arbitrary regular
complex function of the parameters, while $r_{0}(u,\zeta ,\overline{\zeta }%
)\ $is a real function that determines the arbitrarily origin of the affine
parameter along each geodesic. The null tetrad vectors, $l^{a},m^{a},%
\overline{m}^{a},n^{a}\ $are given by%
\begin{eqnarray}
l^{a} &=&\frac{\sqrt{2}}{2P}(1+\zeta \overline{\zeta },\text{ }\zeta +%
\overline{\zeta },\text{ }-i(\zeta -\overline{\zeta }),-1+\zeta \overline{%
\zeta });  \label{tetrad} \\
m^{a} &=&\text{\dh }l^{a}=\frac{\sqrt{2}}{2P}(0,1-\overline{\zeta }^{2},-i(1+%
\overline{\zeta }^{2}),\text{ }2\overline{\zeta }),  \notag \\
\overline{m}^{a} &=&\overline{\text{\dh }}l^{a}=\frac{\sqrt{2}}{2P}%
(0,1-\zeta ^{2},\text{ }i(1+\zeta ^{2}),2\zeta ),  \notag \\
n^{a} &=&\frac{\sqrt{2}}{2P}(1+\zeta \overline{\zeta },-(\zeta +\overline{%
\zeta }),\text{ }i(\zeta -\overline{\zeta }),1-\zeta \overline{\zeta }), 
\notag \\
P &=&1+\zeta \overline{\zeta }.  \label{P}
\end{eqnarray}

\textbf{Aside: }We note that Eq.(\ref{ngc}) has the alternative
interpretation as a coordinate transformation between the $x^{a}\ $and the $%
(u,\zeta ,\overline{\zeta },r)$.

The optical parameters\cite{NP}\cite{Scholarpedia} associated with the
congruence, i.e., the complex divergence $\rho ,$ the complex shear $\sigma
\ $and the twist $\Sigma ,$ are given, after a rather lengthy calculation,
by 
\begin{eqnarray}
\rho &=&\frac{i\Sigma -(r-r_{0}^{\ast })}{r^{2}+\Sigma ^{2}-\overline{\sigma 
}_{0}\sigma _{0}}  \label{optical} \\
\sigma &=&\frac{\sigma _{0}}{(r-r_{0}^{\ast })^{2}+\Sigma ^{2}-\overline{%
\sigma }_{0}\sigma _{0}}  \notag
\end{eqnarray}%
with 
\begin{eqnarray}
\sigma _{0} &=&\text{\dh }L+LL,_{u}  \label{sigmao} \\
2i\Sigma &=&\text{\dh }\overline{L}+L\overline{L},_{u}-\text{ }\overline{%
\text{\dh }}L-\overline{L}L,_{u}\text{ .}  \label{SIGMA}
\end{eqnarray}%
The arbitrary function $r_{0}\ $has first been chosen as 
\begin{equation*}
r_{0}=-\frac{1}{2}(\text{\dh }\overline{L}+L\overline{L},_{u}+\overline{%
\text{\dh }}L+\overline{L}L,_{u})
\end{equation*}%
and then a new arbitrary $r_{0}^{\ast }(u,\zeta ,\overline{\zeta })$ chosen
again as the origin for the affine parameter $r$.

We thus see that the optical parameters are determined by the choice of $%
L(u,\zeta ,\overline{\zeta }).$

Our main interest lies in the class of \textit{regular} null geodesic
congruences with a vanishing shear. This condition is achieved by imposing
the shear-free condition on $L,$ i.e., that $L$ must satisfy the
differential condition.\cite{LR} 
\begin{equation}
\text{\dh }L+LL,_{u}=0  \label{Shearfree}
\end{equation}%
and with regular solutions. \ This procedure has been well documented\cite%
{KandN},\cite{LR} in the literature and we only give the solution.

Changing the independent variable $u\ $\ to $\tau \ $via the function 
\begin{equation}
u=G^{\ast }(\tau ,\zeta ,\overline{\zeta }),  \label{u=G}
\end{equation}%
with inverse,%
\begin{equation}
\tau =T(u,\zeta ,\overline{\zeta }),  \label{tau}
\end{equation}%
we have the solution of Eq.(\ref{Shearfree}), given parametrically, as 
\begin{eqnarray}
L(u,\zeta ,\overline{\zeta }) &=&\text{\dh }_{(\tau )}G^{\ast }(\tau ,\zeta ,%
\overline{\zeta })|_{\tau =T(u,\zeta ,\overline{\zeta })}  \label{L} \\
u &=&G^{\ast }(\tau ,\zeta ,\overline{\zeta })=\xi ^{a}(\tau )l_{a}(\zeta ,%
\overline{\zeta }).  \label{G}
\end{eqnarray}

The operator\ \dh $_{(\tau )}$ means \dh\ holding $\tau \ $constant. \ The $%
z^{a}=\xi ^{a}(\tau )\ $determines\ a complex$\ $world-line, (parametrized
by the complex $\tau ),\ $in a$\ $complex$\ $four-dimensional space that can
be identified with complex Minkowski space - a special case of $\mathcal{H-}$%
space. \ 

When the complex world-line is chosen as a real world-line in real Minkowski
space, i.e., $z^{a}=>x^{a}=$ $\xi _{R}^{a}(t),\ $with $t$ chosen so that the
velocity vector, $v^{a}=\dot{\xi}_{R}^{a}\ ,$ is time-like with norm 
\TEXTsymbol{\vert}v$^{a}$\TEXTsymbol{\vert}$=1,\ $we have the family of null
geodesics given by the generators of the light-cones with apex on the
world-line $x^{a}=\xi _{R}^{a}(t).\ \ $Explicitly, Eq.(\ref{ngc})\ becomes
(after appropriately adjusting the $r_{0}^{\ast }$), our LCs,%
\begin{equation*}
x^{a}=\xi ^{a}(u)+rl^{a}(\zeta ,\overline{\zeta }).
\end{equation*}

If we chose $\xi ^{a}(\tau )\ $as a complex curve, (again with the velocity
normalization, \TEXTsymbol{\vert}v$^{a}$\TEXTsymbol{\vert}$=1$) the
construction of the congruence is a bit more complicated. In the equations, (%
\ref{L})\ and (\ref{G}), $u,\ \tau ,G^{\ast }$\ are complex while in
principle we need $u$ to be real. \ The problem is handled as follows: $L\ $%
and $\overline{L}$ are first constructed by 
\begin{eqnarray}
L &=&\eth _{(\tau )}G(\tau ,\zeta ,\overline{\zeta })=\xi ^{a}(\tau
)m_{a}(\zeta ,\overline{\zeta }),  \label{L?} \\
\overline{L} &=&\overline{\xi }^{a}(\overline{\tau })\overline{m}_{a}(\zeta ,%
\overline{\zeta })  \notag
\end{eqnarray}

In Eq.(\ref{u=G}) we replace $\tau \ $by $t+i\Lambda \ $and decompose the
complex $G$\ into its real and imaginary parts,%
\begin{equation*}
G^{\ast }(t+i\Lambda ,\zeta ,\overline{\zeta })=G_{R}^{\ast }(t,\Lambda
,\zeta ,\overline{\zeta })+iG_{I}^{\ast }(t,\Lambda ,\zeta ,\overline{\zeta }%
).
\end{equation*}%
By setting $G_{I}^{\ast }(t,\Lambda ,\zeta ,\overline{\zeta })=0\ $we$\ $%
determine $\Lambda =\Lambda (t,\zeta ,\overline{\zeta })\ $and
simultaneously make $u\ $real, $u_{R}=G_{R}^{\ast }(t,\Lambda ,\zeta ,%
\overline{\zeta }).\ $Its inverse

\begin{equation*}
t=T_{R}(u_{R},\zeta ,\overline{\zeta })
\end{equation*}%
with $\Lambda (t,\zeta ,\overline{\zeta })\ $allow us to express $L,%
\overline{L},\ $Eq.(\ref{L?}),\ as functions of ($u_{R},\zeta ,\overline{%
\zeta }$). \ We drop use of the subscript ($R$) in the $u.$

The null geodesic congruence, Eq.(\ref{ngc}), constructed with these $L\ $%
and $\overline{L}\ $are the ACLCs. \ 

The twist, Eq.(\ref{SIGMA}), of these congruences is determined by the
imaginary part of the world-line, $\xi _{I}^{a}(\tau ),\ [\xi ^{a}=\xi
_{R}^{a}+i\xi _{I}^{a}]\ $via 
\begin{equation}
\Sigma =\xi _{I}^{a}(\tau )(n_{a}-l_{a}).  \label{twist}
\end{equation}

\textbf{Note: }\ It is important that we first construct the $L\ $and $%
\overline{L}\ $by taking the $\eth _{(\tau )}\ $and $\overline{\eth }_{(\tau
)}\ $derivatives before choosing $u\ $to be real.

\textbf{Aside: }If instead of $\bar{L}\ $we had used $\widetilde{L}=\xi
^{a}(\tau )\overline{m}_{a}(\zeta ,\overline{\zeta })\ $with the complex $u\ 
$from Eq.(\ref{tau})\ the congruence would have been complex Minkowski space
light-cones - i.e., the reason for referring those constructed with $\bar{L}%
\ $as \textit{almost complex cones}.

\textbf{Remark: }We mention, for later use, that the complex world-line $%
z^{a}=\xi ^{a}(\tau )\ $will be uniquely chosen, \textit{by definition}, as
both the complex center of mass, (center of mass +i angular momentum) and 
\textit{complex center of charge, (electric dipole+i magnetic dipole)
world-line} on which both vanish. That both "centers" vanish on the same
world-line is a special case of the more general situation.

\subsubsection{An Alternative Means of Construction}

For the insight that it gives and for use in the following section in the
determination of \textit{asymptotically shear-free congruences,} we describe
an alternative method of construction of these congruences.

Starting with the family of null geodesics from the null cones with apex on
the spatial origin

\begin{equation}
x^{a}=ut^{a}+rl^{a}(\zeta ,\overline{\zeta }),  \label{simple congruence}
\end{equation}%
(augmented by $l^{a},m^{a},\overline{m}^{a},n^{a}$ \ from Eq.(\ref{tetrad}))
we can, very roughly or intuitively, define Penrose's Null Infinity, $%
\mathfrak{I}^{+},\ $as all the points, ($u,\zeta ,\overline{\zeta }$)
obtained by taking the limit, $r=>\infty .\ \mathfrak{I}^{+}\ $\ becomes the
null surface at null infinity - with the structure of $\mathcal{R}$x$%
\mathcal{S}^{2}.\ \ $It is obviously coordinatized by the ($u,\zeta ,%
\overline{\zeta }$). (The tetrad, with these coordinates, ($u,\zeta ,%
\overline{\zeta }$), are a special case of a Bondi system.) It turns out\cite%
{Kent} that the forward light-cone from any\textit{\ interior} \textit{%
space-time }point $x^{a}\ $intersects $\mathfrak{I}^{+}\ $on the cut or
slice, $\mathcal{S}^{2},$ of $\mathfrak{I}^{+}\ $given by 
\begin{equation}
u=x^{a}l_{a}(\zeta ,\overline{\zeta })  \label{focus}
\end{equation}%
so that the light-cones of a space-time world-line, $x^{a}=\xi _{R}^{a}(\tau
),$ yields a one-parameter family of slicings of $\mathfrak{I}^{+},$ 
\begin{equation}
u=G^{\ast }(\tau ,\zeta ,\overline{\zeta })=\xi _{R}^{a}(\tau )l_{a}(\zeta ,%
\overline{\zeta }).  \label{slicings}
\end{equation}

The null vectors, $l^{^{\ast }a},$ tangent to the geodesic congruence coming
from the interior, that are normal to the slicings are given by%
\begin{equation}
l^{^{\ast }a}=l^{a}+L\overline{m}^{a}+\bar{L}m^{a}+L\bar{L}n^{a}
\label{l >.l*}
\end{equation}%
with $L\ $=\ \dh $G.\ $The condition for the congruence to be shear-free is
again, Eq.(\ref{Shearfree}), \dh $L$+$LL,_{u}=0\ $which is satisfied by Eq.(%
\ref{slicings}).$\ $We are thus back to the previous discussion and getting
close to the discussion of the next section.

If the world-line is real then we can replace the $\tau \ $by $t$, if it is
complex, again we construct $L\ $=\ \dh $G^{\ast }\ $as done earlier and
evaluate it for real $u.$

\section{Asymptotically Flat Space-Times}

The study of solutions and properties of \ the asymptotically flat
Einstein-Maxwell equations is a large subject with a great deal of
literature. We will need only a small fraction of this material. \ Rather
than rederiving what we do need, we will largely take from this literature -
mainly from Newman-Penrose, (in Scholarpedia) and Adamo-Newman (in Living
Reviews) - often making use of the NP formalism\cite{Scholarpedia}\cite{NP}%
\cite{NU}

A basic tool in these studies was the introduction, by Bondi, of null
surfaces to be used as part of the asymptotic coordinate system. A
one-parameter family of null surfaces, $\mathfrak{B}_{u}$ labeled by $u,$
was introduced. A two-parameter family of null geodesics, the generators or
geodesics of each surface, are each labeled by sphere coordinates $(\theta
,\phi )$ or equivalently (used here) by complex stereographic coordinates ($%
\zeta ,\overline{\zeta }$), where $\zeta =e^{i\phi }\cot (\frac{\theta }{2}%
). $ The `length' along the geodesics is given by the affine parameter, $r$.
Again (as in the previous section), roughly or intuitively, the future null
boundary of space-time, i.e., Penrose's Null Infinity, $\mathfrak{I}^{+},\ $%
is defined by points ($u$,$\zeta ,\overline{\zeta }$) taken in the limit, $%
r=>\infty .$ The 'boundary', $\mathfrak{I}^{+},\ ($\textit{which can be
mathematically more formally} \textit{defined}), is a null surface, $%
\mathcal{S}^{2}$x$\mathcal{R},\ $coordinatized by ($u$,$\zeta ,\overline{%
\zeta }$) with $u$ as the intersection points of $\mathfrak{I}^{+}\ $with
the Bondi null surfaces $\mathfrak{B}_{u}.\ $The generators of $\mathfrak{I}%
^{+},\ ($the $\mathcal{S}^{2}\ $\ part$)\ $are labeled by the stereographic
coordinates ($\zeta ,\overline{\zeta }$) - and have the same labels as the
generators of $\mathfrak{B}_{u}\ $that they intersect.

\textbf{Aside: }The full set of coordinates, ($u$,$\zeta ,\overline{\zeta }%
,r $), called Bondi coordinates are not unique - there being an entire
group, the BMS group of transformations\cite{LR}, connecting the different
members. This lack of uniqueness does not now play an important role for us
- though that is likely to change in the future.

In addition to Bondi coordinates, a Bondi system, at and near $\mathfrak{I}%
^{+}$ also contains a null tetrad, ($l_{B}^{a},m_{B}^{a},\overline{m}%
_{B}^{a},n_{B}^{a}$)$.\ $The $l_{B}^{a}\ $are tangent vectors to the
geodesics of $\mathfrak{B}_{u},\ $the $n_{B}^{a},$ $\ $are tangent vectors
of the generators of $\mathfrak{I}^{+},$\ $(m_{B}^{a},\overline{m}_{B}^{a})\ 
$are tangent vectors to the $u=const\ $slices of $\mathfrak{I}^{+}.\ $\ The $%
\ (n_{B}^{a},m_{B}^{a},\overline{m}_{B}^{a})\ $are parallel \ propagated
down the generators of $\mathfrak{B}_{u}\ $to the interior.

The points of $\mathfrak{I}^{+}\ $with constant value of $u\ $are referred
to as Bondi slices or Bondi cuts; any arbitrary cross-section\ or family of
cross-section of $\mathfrak{I}^{+},\ $i.e.,$\ u=K(\zeta ,\overline{\zeta })\ 
$or $u=F(s,\zeta ,\overline{\zeta })\ $\ are called slices or cuts. \ Much
of our effort will be devoted to finding, studying and giving applications
to certain preferred slicings (\textit{asymptotically shear-free)} - that
are very different from a Bondi slicing. \ An important fact is that the
family of null geodesics of the surfaces $\mathfrak{B}_{u},\ $in general, 
\textit{are not (asymptotically) shear-free. }Their shear is given by\textit{%
\ }%
\begin{equation*}
\sigma =\frac{\sigma ^{0}(u,\zeta ,\overline{\zeta })}{r^{2}}+O(r^{-4}),
\end{equation*}%
with $\sigma ^{0}(u,\zeta ,\overline{\zeta })\ $referred to as the
asymptotic shear. It plays the role of arbitrary radiation data. ($\sigma
^{0},_{u}\ $is referred to as the Bondi news function.) Our task is to find
slicings of $\mathfrak{I}^{+}$ so that the \textit{normal null congruences} 
\textit{(normal to the slicing) }are asymptotically shear-free. \ 

Since this problem has been described and solved in the literature\cite{LR},
we give the solution with just a brief explanation.

The Sachs theorem, which describes how $\sigma ^{0}(u,\zeta ,\overline{\zeta 
})\ $transforms under a BMS super-translations, i.e., under $u^{\prime
}=u-\alpha (\zeta ,\overline{\zeta }),\ $states that 
\begin{equation*}
\sigma ^{0\ \prime }(u^{\prime },\zeta ,\overline{\zeta })=\sigma
^{0}(u,\zeta ,\overline{\zeta })-\eth ^{2}\alpha .
\end{equation*}%
Setting the new shear to zero, $\sigma ^{0\ \prime }\ =0$ when $u^{\prime
}==0,\ ($i.e., at $u=\alpha (\zeta ,\overline{\zeta })),\ $leads, with $%
\alpha $ replaced by $G(\zeta ,\overline{\zeta }),$ to 
\begin{equation}
\eth ^{2}G=\sigma ^{0}(G,\zeta ,\overline{\zeta }),  \label{GCEq}
\end{equation}%
the \ so-called "good-cut Equation".

Solutions to Eq.(\ref{GCEq}) have been shown\cite{LR}\cite{Hansen} to depend
on four arbitrary complex numbers, $z^{a},\ \ $which in turn define a
four-complex dimensional space, referred to as $\mathcal{H}-$space. \
Imposing coordinate conditions on the choices of these \ coordinates, the
solution can always be written as%
\begin{equation}
u=G(z^{a},\zeta ,\overline{\zeta })=z^{a}l_{a}(\zeta ,\overline{\zeta }%
)+H_{l\eqslantgtr 2}(z^{a},\zeta ,\overline{\zeta })  \label{solution w z}
\end{equation}%
with $H_{l\eqslantgtr 2}(z^{a},\zeta ,\overline{\zeta })\ $expandable in
spherical harmonics $\ l\eqslantgtr 2$.

\ \textbf{Aside: }We mention without further discussion that $\mathcal{H}-$%
space has a variety of interesting properties\cite{Hansen}\cite{LR}: it
possesses a complex holomorphic metric, it is Ricci flat and is anti-self
dual.

By choosing an arbitrary "world-line", $z^{a}=\xi ^{a}(\tau ),\ $we have a
one-complex$\ $parameter family of cuts of $\mathfrak{I}^{+},\ ($%
complexified in general$).\ $

\begin{equation}
u=G(\xi ^{a}(\tau ),\zeta ,\overline{\zeta })\equiv G^{\ast }(\tau ,\zeta ,%
\overline{\zeta })=\xi ^{a}(\tau )l_{a}(\zeta ,\overline{\zeta }%
)+H_{l\eqslantgtr 2}(\xi ^{a}(\tau ),\zeta ,\overline{\zeta }).  \label{G*}
\end{equation}%
Using the freedom of reparametrization, $\tau ^{\ast }=F(\tau ),\ $we make $%
\dot{\xi}^{a}\dot{\xi}_{a}\approx 1\ $via a slow motion approximation, so
that $\xi ^{0}\approx \tau ,\ \dot{\xi}^{i}\dot{\xi}_{i}\approx 0.$This
leads to

\begin{equation}
u=G=\frac{\tau }{\sqrt{2}}-\frac{1}{2}\xi ^{i}(\tau )Y_{1i}^{0}(\zeta ,%
\overline{\zeta })+\xi ^{ij}(\xi ^{a}(\tau ))Y_{ij}^{2}(\zeta ,\overline{%
\zeta })+..  \label{G**}
\end{equation}

In the following section $\xi ^{a}(\tau )$ will be chosen in two separate
ways: first $\xi ^{a}(\tau )\ $is taken - by definition - as the unique
complex center of mass world-line, 
\begin{equation}
z^{a}=\xi _{CofM}^{a}(\tau ),  \label{CofM}
\end{equation}%
still to be determined.

\textbf{Remark: }As mentioned earlier, we specialize by assuming that the
complex center of charge world-line coincides with the complex center of
mass. This is not necessary but is a restriction on the class of solutions.

The second choice for $\xi ^{a}$, again - by definition - is 
\begin{eqnarray}
z^{a} &=&\tau ^{\ast }\delta _{0}^{a},  \label{static frame} \\
u &=&G^{\ast }(\tau ^{\ast },\zeta ,\overline{\zeta })=\frac{\tau ^{\ast }}{%
\sqrt{2}}+\xi ^{ij}(\tau ^{\ast })Y_{ij}^{2}(\zeta ,\overline{\zeta })+..
\label{u<>tau*}
\end{eqnarray}%
$\ $yielding the "static-frame"$.\ \ $Note that the these "static" slicing
differ from Bondi slicing by $l\eqslantgtr 2\ $harmonics$\ $and are very
close to Lorentzian-looking slicings.

The associated asymptotically shear-free congruences that are normal to the
slicings are determined by%
\begin{equation*}
l^{^{\ast }a}=l_{B}^{a}+L\overline{m}_{B}^{a}+\bar{L}m_{B}^{a}+L\bar{L}%
n_{B}^{a}
\end{equation*}%
with%
\begin{eqnarray}
L\ &=&\ \mathbf{\eth }_{(\tau )}G^{\ast },\   \label{L&Lbar} \\
\overline{L}\ &=&\ \overline{\mathbf{\eth }}_{(\tau )}\overline{G}^{\ast }.\ 
\notag
\end{eqnarray}%
The $L\ \ $automatically, from its construction, satisfies, parametrically,
the generalization of Eq.(\ref{Shearfree}), namely 
\begin{eqnarray*}
\eth L+LL,_{u} &=&\sigma ^{0}(u,\zeta ,\overline{\zeta }) \\
u &=&G^{\ast }(\tau ,\zeta ,\overline{\zeta }).
\end{eqnarray*}

The asymptotic twist $\Sigma \ $is almost the same as in Eq.(\ref{twist})

\begin{equation}
\Sigma =\xi _{I}^{a}(\tau )(n_{a}-l_{a})+higher\ harmonics  \label{Sigma}
\end{equation}

If the world-line and $G^{\ast }\ $are real then we can replace the $\tau \ $%
by $t$, if complex, again we must construct $L\ =\mathbf{\eth }_{(\tau
)}G^{\ast }\ $and then evaluate it for real $u,\ $as in the previous section.

We are back to virtually the same results and discussion as that of the
previous section in the flat-space "Alternative Means of Construction" .

Our congruences are then the geodesics of the ALCs, i.e., they are
asymptotically shear-free and they are labeled by points in a four complex
dimensional space - $\mathcal{H}$-space. \ In the special case of passing to
the limit of flat space, the congruences do then becomes those of LCs with
the labeling remaining.

\section{Application}

Much of the material of this section - with detailed lengthy derivations -
have appeared earlier\cite{NP}\cite{NU}\cite{LR} Here, in the context of our
ALCs, we will simply describe these results, with some explanations but
little in the way of derivation.

In this section 'prime'\ will denote the $u$-derivative.

We start with an asymptotically flat Einstein-Maxwell solution described in
the neighborhood of null infinity, in a Bondi coordinate system with a Bondi
tetrad. \ Using NP\cite{NP} notation, the Weyl and Maxwell tensors have the
asymptotic (peeling) behavior,

\begin{eqnarray}
\Psi _{0} &=&\Psi _{0}^{0}r^{-5}+O(r^{-6}),  \label{Weyl} \\
\Psi _{1} &=&\Psi _{1}^{0}r^{-4}+O(r^{-5}),  \notag \\
\Psi _{2} &=&\Psi _{2}^{0}r^{-3}+O(r^{-4}),  \notag \\
\Psi _{3} &=&\Psi _{3}^{0}r^{-2}+O(r^{-3}),  \notag \\
\Psi _{4} &=&\Psi _{4}^{0}r^{-1}+O(r^{-2}).  \notag
\end{eqnarray}
\begin{eqnarray}
\phi _{0} &=&\phi _{0}^{0}r^{-3}+O(r^{-4}),  \label{Maxwell} \\
\phi _{1} &=&\phi _{1}^{0}r^{-2}+O(r^{-3}),  \notag \\
\phi _{2} &=&\phi _{2}^{0}r^{-1}+O(r^{-2}).  \notag
\end{eqnarray}

The $\Psi _{n}^{0}\ \ $and $\phi _{n}^{0}\ ,\ $live on$\ \mathfrak{I}^{+},\ $%
i.e., are functions of ($u$,$\zeta ,\overline{\zeta }$). \ They satisfy the
asymptotic Bianchi Identities and asymptotic Maxwell equations,%
\begin{eqnarray}
\Psi _{2}^{0\,\prime } &=&-\text{\dh }\Psi _{3}^{0\,}+\sigma ^{0}\Psi
_{4}^{0\,}+k\phi _{2}^{0}\overline{\phi }_{2}^{0},  \label{AsyBI1} \\
\Psi _{1}^{0\,\prime } &=&-\text{\dh }\Psi _{2}^{0\,}+2\sigma ^{0}\Psi
_{3}^{0\,}+2k\phi _{1}^{0}\overline{\phi }_{2}^{0},  \label{AsyBI2} \\
\Psi _{0}^{0\,\prime } &=&-\text{\dh }\Psi _{1}^{0\,}+3\sigma ^{0}\Psi
_{2}^{0\,}+3k\phi _{0}^{0}\overline{\phi }_{2}^{0},  \label{AsyBI3} \\
k &=&2Gc^{-4},
\end{eqnarray}%
\begin{eqnarray}
\phi _{1}^{0\,\prime } &=&-\text{\dh }\phi _{2}^{0},  \label{MaxI} \\
\phi _{0}^{0\,\prime } &=&-\text{\dh }\phi _{1}^{0}+\sigma ^{0}\phi _{2}^{0},
\label{MaxII}
\end{eqnarray}%
the prime denoting the $u$-derivative. $\sigma ^{0}$($u$,$\zeta ,\overline{%
\zeta }$)\ is the asymptotic shear, the free data. From the field equation
we have that%
\begin{eqnarray}
\Psi _{3}^{0} &=&\text{\dh }(\overline{\sigma }^{0})^{\prime }\ ,
\label{PSI_3,4} \\
\Psi _{4}^{0} &=&-(\overline{\sigma }^{0})^{\prime \prime }.  \notag
\end{eqnarray}%
It is very convenient to introduce, instead of the $\Psi _{2}^{0\,},\ $the
mass aspect ${\large \Psi ,}$ (which is real from the field equations) by%
\begin{equation}
\Psi =\overline{\Psi }\equiv \Psi _{2}^{0\,}+\eth ^{2}\overline{\sigma }%
^{0}+\sigma ^{0}(\overline{\sigma }^{0})^{\prime },  \label{MassAspect}
\end{equation}

Bondi defines the asymptotic mass, $M_{B},$ and (R.Sachs) the 3-momentum, $%
P_{B}^{i}\ \ $as the$\ l=0\ $\&$\ l=1\ $harmonic coefficients of $\Psi .\ $%
Specifically,

\textbf{Definition 1 \ Identification of Physical Quantities:}%
\begin{eqnarray}
\Psi &=&\Psi ^{0}+\Psi ^{i}Y_{1i}^{0}+\Psi ^{ij}Y_{2ij}^{0}+...
\label{DEF.1} \\
\Psi ^{0} &=&-\frac{2\sqrt{2}G}{c^{2}}M_{B}  \label{mass} \\
\Psi ^{i} &=&-\frac{6G}{c^{3}}P_{B}^{i}  \label{momentum}
\end{eqnarray}

By rewriting Eq.(\ref{AsyBI1}), replacing the $\Psi _{2}^{0\,}$ by $\Psi $
via Eq.(\ref{MassAspect}), we have

\begin{equation*}
\Psi ^{\prime }\text{ }=\text{ }(\sigma ^{0})^{\prime }(\overline{\sigma }%
^{0})^{\prime }+\ k\phi _{2}^{0}\overline{\phi }_{2}^{0},
\end{equation*}%
and one immediately has the Bondi mass/energy loss theorem - which we return
to later:%
\begin{equation}
M_{B}^{\prime }=-\frac{c^{2}}{2\sqrt{2}G}\int ((\sigma ^{0})^{\prime }(%
\overline{\sigma }^{0})^{\prime }+k\phi _{2}^{0}\overline{\phi }%
_{2}^{0})d^{2}S\ \preceq 0.  \label{BondiTheorem}
\end{equation}

In addition to the Bondi/Sach energy-momentum we \textit{define the complex
center of mass} by the $l=1\ $(complex) spherical harmonic component of $%
\Psi _{1}^{0\,}.\ $This definition, which came originally from linear
theory, is now justified by the results that it leads to:

\textbf{Definition 2 \ Complex Center of Mass}

\begin{equation}
\Psi _{1}^{0}=-6\sqrt{2}Gc^{-2}(D_{(mass)}^{i}+ic^{-1}J^{i})Y_{1i}^{1}+....
\label{DEF.2}
\end{equation}%
with $D_{(mass)}^{i}\ $the mass dipole and $J^{i},\ $the total angular
momentum, as seen at null infinity. \ 

\textbf{\ \ }Our physical identification (standard) for the complex E\&M
dipole, (electric and magnetic dipoles) as the $l=1$ harmonic component of $%
\phi _{0}^{0}\ $and electric charge $q\ $are:

\textbf{Definition\ \ 3 \ \ Complex E\&M Dipole\ and Charge} with the $Q\ $s
representing known quadrupole terms and $q$ the Coulomb charge.%
\begin{eqnarray}
(D_{Elec}^{i}+iD_{Mag}) &=&q\xi ^{i}  \label{DEF.3} \\
\phi _{0}^{0} &=&2q\xi ^{i}Y_{1i}^{1}+Q_{0}...  \notag \\
\phi _{1}^{0} &=&q+\sqrt{2}q\xi ^{i\ \prime }Y_{1i}^{0}+Q_{1}+...  \notag \\
\phi _{2}^{0} &=&-2q\xi ^{i\ \prime \prime }Y_{1i}^{-1}+Q_{2}+...,  \notag
\end{eqnarray}

We have made, as mentioned earlier, a simplifying assumption here, namely
that the complex center of charge coincides with the complex center of mass.
This is not necessary but is a chosen special case.\qquad

Our main interests lie in the components $\Psi _{1}^{0\,}\ $and$\ \Psi ,\ $%
with their physical identifications and their evolution equations, (\ref%
{AsyBI1}) and (\ref{AsyBI2}).

Our modus operandi is now to consider both the tetrad and coordinate
transformations from the Bondi tetrad and coordinates to the coordinates and
associated tetrad of a ALC with (for the moment) an arbitrary complex
world-line, $z^{a}=\xi ^{a}(\tau ).$\ The transform of the tetrad and Weyl
tensor components are

\begin{eqnarray}
l^{\ast a} &=&l_{B}^{a}+b\overline{m}_{B}^{a}+\overline{b}%
m_{B}^{a}+0(r^{-2}),  \label{null rot II} \\
m^{\ast a} &=&m_{B}^{a}+bn_{B}^{a},  \notag \\
n^{\ast a} &=&n_{B}^{a},  \notag \\
b &=&-\frac{L}{r}+0(r^{-2}).  \notag
\end{eqnarray}%
and

\begin{eqnarray}
\Psi _{0}^{\ast 0} &=&\Psi _{0}^{0}-4L\Psi _{1}^{0}+6L^{2}\Psi
_{2}^{0}-4L^{3}\Psi _{3}^{0}+L^{4}\Psi _{4}^{0},  \label{0} \\
\Psi _{1}^{\ast 0} &=&\Psi _{1}^{0}-3L\Psi _{2}^{0}+3L^{2}\Psi
_{3}^{0}-L^{3}\Psi _{4}^{0},  \label{1} \\
\Psi _{2}^{\ast 0} &=&\Psi _{2}^{0}-2L\Psi _{3}^{0}+L^{2}\Psi _{4}^{0},
\label{2} \\
\Psi _{3}^{\ast 0} &=&\Psi _{3}^{0}-L\Psi _{4}^{0},  \label{3} \\
\Psi _{4}^{\ast 0} &=&\Psi _{4}^{0}.  \label{4}
\end{eqnarray}

The $L\ $and its complex conjugate, $\overline{L},$ are determined by Eq.(%
\ref{L&Lbar}) with the coordinate transformation from $u$ to $\tau \ $given\
by, Eq.(\ref{G**}),

\begin{equation}
u=\frac{\tau }{\sqrt{2}}-\frac{1}{2}\xi ^{i}(\tau )Y_{1i}^{0}(\zeta ,%
\overline{\zeta })+\xi ^{ij}(\xi ^{a}(\tau ))Y_{ij}^{2}(\zeta ,\overline{%
\zeta })+..  \label{cuts II}
\end{equation}

The world-line $\xi _{CofM}^{a}=\ $($\tau ,\xi ^{i}(\tau )$) is now
determined by setting to \textit{zero the three components of the }$l=1$%
\textit{\ coefficients} of $\Psi _{1}^{\ast 0}\ $in Eq.(\ref{1}). \ Actually
rather than doing that we reverse the process, using $\Psi _{1}^{\ast 0i}=0,$
and express the original Bondi $\Psi _{1}^{0}\ $in\ terms\ of\ the $\xi
_{CofM}^{a}(\tau ).$ Finally, after considerable effort, with Taylor and
Clebsch-Gordon products and expansions, we have the $l=1$ harmonic
coefficient of$\ $the Bondi $\Psi _{1}^{0}$\cite{LR},

\begin{eqnarray*}
\Psi _{1}^{0i} &=&-\frac{6\sqrt{2}G}{c^{2}}M_{B}\xi _{CofM}^{i}+i\frac{6%
\sqrt{2}G}{c^{3}}P_{B}^{k}\xi _{CofM}^{j}\epsilon _{kji}-\frac{576G}{5c^{3}}%
P_{B}^{k}\xi ^{ik}+i\frac{6912\sqrt{2}}{5}\xi ^{lj}\overline{\xi }%
^{lk}\epsilon _{jki} \\
&&-i\frac{2\sqrt{2}G}{c^{6}}q^{2}\xi _{CofM}^{k}\overline{\xi }^{j\prime
\prime }\epsilon _{kji}-\frac{48G}{5c^{6}}q^{2}\xi ^{ji}\overline{\xi }%
^{j\prime \prime }-\frac{4G}{5c^{7}}q^{2}\xi _{CofM}^{j}\overline{Q}%
_{C}^{ij\prime \prime \prime }-i\frac{16\sqrt{2}G}{5c^{7}}q\xi ^{lj}%
\overline{Q}_{C}^{lk\prime \prime \prime }\epsilon _{jki}.
\end{eqnarray*}

The Bondi-Sachs mass-momentum, \textbf{Definition 1,} has already been used.
Prime indicates $u\ $derivative. \ Assuming that the quadrupole interactions
(E\&M and gravitational) and the high time derivatives are small, we are
left with

\begin{equation}
\Psi _{1}^{0i}=-\frac{6\sqrt{2}G}{c^{2}}M_{B}\xi _{CofM}^{i}+i\frac{6\sqrt{2}%
G}{c^{3}}P_{B}^{k}\xi _{CofM}^{j}\epsilon _{kji}.  \label{PSIi1}
\end{equation}

Finally using 
\begin{equation}
\xi _{CofM}^{i}=\xi _{R}^{i}+i\xi _{I}^{i},  \label{real&complex}
\end{equation}%
and comparing Eq.(\ref{PSIi1}) with our \textbf{definition 2,} $\Psi
_{1}^{0i}=-6\sqrt{2}Gc^{-2}(D_{(mass)}^{i}+ic^{-1}J^{i}),\ $we obtain our

\textbf{Result:1 - Dipole and Angular momentum}

\begin{eqnarray}
D_{(mass)}^{i} &=&M_{B}\xi _{R}^{i}-c^{-1}P_{B}^{k}\xi _{I}^{j}\ \epsilon
_{jki}+...,  \label{mass dipole} \\
J^{i} &=&cM_{B}\xi _{I}^{i}+P_{B}^{k}\xi _{R}^{j}\epsilon _{jki}+....
\label{ang mom}
\end{eqnarray}%
or 
\begin{eqnarray}
\overrightarrow{D}_{(mass)} &=&M_{B}\overrightarrow{r}+c^{-2}M_{B}^{-1}%
\overrightarrow{P}_{B\ }\mathrm{x}\overrightarrow{S}.  \label{D} \\
\overrightarrow{r} &=&\xi _{R}^{i}=(\xi _{R}^{1},\xi _{R}^{2},\xi _{R}^{3}),
\\
\overrightarrow{S} &=&cM_{B}\xi _{I}^{j}=cM_{B}(\xi _{I}^{1},\xi
_{I}^{2},\xi _{I}^{3}), \\
\overrightarrow{J} &=&\overrightarrow{S}+\overrightarrow{r}\mathrm{x}%
\overrightarrow{P}.
\end{eqnarray}

The mass dipole is the usual term plus a 2nd term that is part of the
standard relativist angular momentum tensor\cite{AngMom}. We find for the
angular momentum a spin term $\overrightarrow{S}$ and the standard $%
\overrightarrow{r}\mathrm{x}\overrightarrow{P}\ $orbital angular momentum
term. \ 

\textbf{REMARK }Notice that once we have the \textit{definition} of the
complex center of mass and the complex center of mass world-line, these
results for $\overrightarrow{D}_{(mass)}\ $and $\overrightarrow{J}\ $follow
without any further calculations.

Next, replacing $\Psi _{1}^{0i},$ from Eq.(\ref{PSIi1}), in the evolutionary
Bianchi Identity, Eq.(\ref{AsyBI2}),

\begin{equation}
\Psi _{1}^{0\,\prime }=-\eth \Psi _{2}^{0\,}+2\sigma ^{0}\Psi
_{3}^{0\,}+2k\phi _{1}^{0}\overline{\phi }_{2}^{0}.  \label{BIII}
\end{equation}%
with \textbf{definition} 1 and Eq.(\ref{DEF.3}), we find at \textit{linear
order}, directly from the real part, the$\ $linear$\ $momentum, $P_{B}^{i}.\ 
$Looking at the lowest harmonic order, the $l=1,\ $we see that\ $P_{B}^{i}\ $%
appears in the $\Psi _{2}^{0\,},\ $the $M_{B}\ \dot{\xi}_{R}^{i\ }\ $\
appears in $\dot{\Psi}_{1}^{0\,}\ $and $\frac{2q^{2}}{3c^{3}}\ddot{\xi}%
_{R}^{i\ }\ \ $is in the last term. This leads immediately - \textit{just by
observation} - to:

\textbf{Result: 2 - Kinematic Linear Momentum }%
\begin{eqnarray}
P_{B}^{i} &=&M_{B}\ \xi _{R}^{i\ \prime }-\frac{2q^{2}}{3c^{3}}\xi _{R}^{i\
\prime \prime }+H.O.  \label{P=Mv} \\
H.O. &=&\text{quadrupole and higher order terms.}  \notag
\end{eqnarray}

We have the Abraham-Lorentz-Dirac radiation reaction term appearing with
virtually no derivation, no assumptions, no mass renormalization - just the
starting definitions.

From the imaginary part of the same Bianchi Identity we have the angular
momentum loss equation;

\textbf{Result: 3 - Angular momentum Conservation}%
\begin{equation}
J^{i\prime }=-\frac{2q^{2}}{3c^{3}}\xi _{I}^{i\ \prime \prime }+\frac{2q^{2}%
}{3c^{3}}(\xi _{R}^{j\ \ \prime }\xi _{R}^{k\prime \prime }+\xi _{I}^{k\
\prime }\xi _{I}^{k\ \prime \prime })\epsilon _{kji}+H.O.\text{ }
\end{equation}

\textbf{Note }The first term on the right side can be moved to the left,
which simply changes the definition of $J^{i}\ ,$ 
\begin{equation}
J^{\ast i\prime }=(J^{i}+\frac{2q^{2}}{3c^{3}}\xi _{I}^{i\ \prime })^{\prime
}=\frac{2q^{2}}{3c^{3}}(\xi _{R}^{j\ \prime }\xi _{R}^{k\ \prime \prime
}+\xi _{I}^{k\ \prime }\xi _{I}^{k\prime \prime })\epsilon _{kji},
\label{J*}
\end{equation}%
i.e., it adds a spin dependent term$.\ $

\textbf{Note }\textit{We have the exact Landau \& Lifschitz\cite{LL}
expression} for angular momentum loss in the special case of Eq.(\ref{J*})
when the derivatives of the spin terms $\xi _{I}^{i}\ $are considered to be
zero.

Finally substituting the Bondi-Sachs terms and those of Eq.(\ref{PSI_3,4})
into the first evolutionary Bianchi Identity, Eq.(\ref{AsyBI1}), 
\begin{subequations}
\begin{equation}
\Psi _{2}^{0\,\prime }=-\eth \Psi _{3}^{0\,}+\sigma ^{0}\Psi
_{4}^{0\,}+k\phi _{2}^{0}\overline{\phi }_{2}^{0},  \label{BI1*}
\end{equation}%
we have for the $l=0$\ harmonic coefficient, the (Bondi) mass loss
expression but now including the well known (classical) electromagnetic
energy losses, i.e.,

\textbf{Result: 4 - Energy loss} 
\end{subequations}
\begin{eqnarray}
M_{B}^{\prime } &=&-\frac{G}{5c^{7}}(Q_{Mass}^{jk\prime \prime \prime
}Q_{Mass}^{jk\prime \prime \prime }+Q_{Spin}^{jk\prime \prime \prime
}Q_{Spin}^{jk\prime \prime \prime })-\frac{4q^{2}}{3c^{5}}(\xi _{R}^{i\prime
\prime }\xi _{R}^{i\prime \prime }+\xi _{I}^{i\prime \prime }\xi
_{I}^{i\prime \prime })  \label{mass loss} \\
&&-\frac{4}{45c^{7}}(Q_{E}^{jk\prime \prime \prime }Q_{E}^{jk\prime \prime
\prime }+Q_{M}^{jk\prime \prime \prime }Q_{M}^{jk\prime \prime \prime }).
\end{eqnarray}%
\noindent \noindent \noindent The first term is the standard Bondi
quadrupole mass loss (now including the \textit{spin-quadrupole}
contribution to the loss - maybe new), the second and third terms are the
classical E\&M dipole and quadrupole energy loss - including the correct
numerical factors. Note again that these results are just sitting in the
Bianchi Identities - with no derivation - arising simply from the Ricci
tensor expressed via the Maxwell stress tensor.

The $l=1\ $terms, the momentum loss expression, leads to

\textbf{Result: 5 - Newton's 2nd Law}

\begin{equation}
P_{B}^{i\ \prime }=F_{recoil}^{i}  \label{P'}
\end{equation}%
where $F_{recoil}^{i}\ $is composed of many non-linear radiation terms
involving the time derivatives of the gravitational quadrupole and the E\&M
dipole and quadrupole moments. These terms are known and given\cite{LR} but
not relevant to us now. \ Instead we substitute Eq.(\ref{P=Mv}) into Eq.(\ref%
{P'}) leading to Newton's second law;

\begin{equation}
M_{B}\xi _{R}^{i\ \prime \prime }=F^{i}\equiv M_{B}^{\prime }\xi _{R}^{i\
\prime }+\frac{2q^{2}}{3c^{3}}\xi _{R}^{\ i\ \prime \prime \prime
}+F_{recoil}^{i}.  \label{F=ma}
\end{equation}%
\textbf{Result: 6 - Rocket Force and Radiation Reaction Force}\qquad \qquad
\qquad

We find this surprising - to have exactly the standard rocket mass loss
expression, i.e.,$\ M^{\prime }v^{\prime },\ $and the exact
Abraham-Lorentz-Dirac radiation reaction force term - no mass
renormalization needed.

\subsection{The Last Step}

For our last step, which turns out to be very easy, we must transform our
results from the Bondi system to the "static frame" of our asymptotic
shear-free system, i.e., Eqs.(\ref{static frame}) and (\ref{u<>tau*}). \ The
coordinate transformation which takes us from the Bondi slicings to the
"static frame" 
\begin{equation*}
u=\frac{\tau ^{\ast }}{\sqrt{2}}+\xi ^{ij}(\tau ^{\ast })Y_{ij}^{2}(\zeta ,%
\overline{\zeta })+..
\end{equation*}%
can within our approximations can be considered simply as 
\begin{equation*}
u=\frac{\tau ^{\ast }}{\sqrt{2}}
\end{equation*}%
with the $L$ and $\overline{L}\ $of the tetrad transformation, Eq.(\ref{null
rot II}) considered as vanishing. \ We can thus treat the transformation as
the identity.

All our six results then hold in the "static frame" using the complex $\tau
^{\ast }\ $instead of the $u.\ \ $By forcing $u\ $to be real, as in the
construction of the previous section, our approximations allow us to treat $%
\Lambda =0,\ $and hence $\tau ^{\ast }=t\ $as real. \ The 'prime'
derivatives can then be thought of as simply $t\ $derivatives. The slicing
of real $\mathfrak{I}^{+}\ $\ are given by

\begin{equation*}
u=t+[\xi ^{ij}(t)Y_{ij}^{2}(\zeta ,\overline{\zeta })]_{R}+..,
\end{equation*}%
namely the Bondi slicing, $(u=t)$, but with small higher, $l$ $\eqslantgtr
2, $ harmonic corrections.

Our results are then the (real) standard relations of classical mechanics.

\section{Discussion}

The results of the previous section raise a variety of issues; some - so far
- have been very difficult to resolve, others raise interesting questions
that remain to be answered or even studied.

\subsection{Meaning}

Our prime problem is the following.

We have found sitting in just the asymptotic Einstein-Maxwell equations,
with no additional physical assumptions, (aside from a few definitions), a
large number of the fundamental relations from classical mechanics coupled
with the Maxwell field. These relations, (e.g., radiation reaction or
angular momentum loss), that often involve considerable effort to obtain by
standard procedures, are simply sitting in the Bianchi identities needing
only the few definitions. The simplicity in finding them - to us - is rather
surprising. \ But even more surprising is the fact that they seem to be
basically unintelligible - they are "equations of motion"\ \ that appear to
have nothing what-so-ever to do with space-time points. \ The "motion" takes
place in the rather unphysical complex $\mathcal{H}$-space. The imaginary
values of the coordinates with their dynamics describe spin angular-momentum
behavior. The real parts of the coordinates mimic real space-time and seem
to describe, in $\mathcal{H}$-space, the motion of the center of mass. \
What does this mean.

A question that can be answered is: do any of these $\mathcal{H}$-space
relations (e.g., the coordinates) appear in any aspects of real space-time
and its $\mathcal{I}^{+}?\ $For each\textit{\ real} ALC cut constructed from
the complex cut, Eq.(\ref{G**}),

\begin{equation*}
u=\frac{\tau }{\sqrt{2}}-\frac{1}{2}\xi ^{i}(\tau )Y_{1i}^{0}(\zeta ,%
\overline{\zeta })+\xi ^{ij}(\xi ^{a}(\tau ))Y_{ij}^{2}(\zeta ,\overline{%
\zeta })+..
\end{equation*}%
the coefficients of the $l=0,1\ $harmonics are the real parts of the $%
\mathcal{H}$-space coordinates while the twist of the congruence determines
the imaginary parts. They have the information to form a \textit{virtual
image} of a point. This is analogous but opposite to the case in flat-space
time where the rays from the cut, Eq.(\ref{focus}), 
\begin{equation*}
u=x^{a}l_{a}(\zeta ,\overline{\zeta }),
\end{equation*}%
\textit{do focus back }to the space-time point $x^{a}.\ \ $Unfortunately
this does not seem\ to help clarify the issue of why these \textit{virtual
points} mimic the behavior of real space-time points.

\subsection{The BMS Group}

The role of the BMS group - the group of coordinate transformations between
different Bondi coordinate systems - appears to be changed by the use of ALC
coordinates. The members of the set of ALCs are geometric constructs and, as
geometric objects, are not subject to intrinsic changes due to an arbitrary
BMS transformation. On the other hand the ALCs are \textit{described}\ in
terms of any - but a specific choice of Bondi coordinates - and do undergo
changes when the specific choice is changed. In other words under a BMS
transformation \textit{the description of the ALCs} will be changed. \ The
details of these changes have not yet been worked out.

\subsection{Queries}

1. \ \ Are our results concerning the classical mechanical relationships at
all significant - or are they just a curious coincidence of little
consequence? \ We feel from the clarity and ease by which they sit in the
Bianchi Identities with their associated Lorentzian-like structures (the
ALCs), that they very likely are significant. But what is that significance?
We also remember that almost \textit{every\ mother loves her own child - }so
we remain skeptical. \ 

2. \ \ In either case, can the results or predictions of the angular
momentum loss and spin contributions to gravitational radiation be
considered as meaningful and correct? Or conceivably measured? These
results, though small, appear to be new.

3. We know that the $\mathcal{H}$-space (and conjugate $\overline{\mathcal{H}%
}$-space) both contain complex-holomorphic Ricci-flat metrics with self-dual
(and anti-self-dual ) Weyl tensors. \ What happens to these structures when
we go to the real $u$ and associated real $\mathcal{H}$-space coordinates?
Do they remain - and if so with what structures?

4. \ The easy appearance of the radiation reaction force is both pleasing
and disturbing. \ No heavy breathing nor hard work, no mass renormalization
or further assumptions. \ It is just there. But then what can we say about
the familiar instability - the runaway behavior of the associated motion due
to radiation reaction? \ Is there a mechanism, that we do not see, that
damps the motion. Or are the solutions to the Einstein-Maxwell equations
unstable? We have no answer.

5. How do these results fold in with the attempts to construct a Quantum
Theory of Gravity. If they do not fold in - then why not? They are part of
GR. If they do fold in, what is their role? \ Do we get a Schrodinger-like
Equation for the center of mass motion or a Dirac Equation for the spin? It
appears highly unlikely.

\section{Acknowledgements}

We thank Timothy Adamo for hours of wonderful discussions and collaboration
on an earlier manuscript where many of the present ideas were developed. \
Roger Penrose is owed, almost beyond thanks, for his insight, his
enlightening remarks and his encouragement and support - both recently and
over the years.

\end{document}